# Three-Dimensional Spin Rotations at the Fermi Surface of a Strongly Spin-Orbit Coupled Surface System

P. Höpfner[1], J. Schäfer[1], A. Fleszar[2], J. H. Dil[3,4], B. Slomski[3,4], F. Meier[3,4], C. Loho[1], C. Blumenstein[1]
L. Patthey[3], W. Hanke[2], and R. Claessen[1]

[1]*Physikalisches Institut, Universität Würzburg, 97074 Würzburg, Germany*
[2]*Institut für Theoretische Physik und Astrophysik, Universität Würzburg, 97074 Würzburg, Germany*
[3]*Swiss Light Source, Paul-Scherrer-Institut, 5232 Villigen, Switzerland*
[4]*Physik-Institut, Universität Zürich, 8057 Zürich, Switzerland*

The spin texture of the metallic two-dimensional electron system ($\sqrt{3}\times\sqrt{3}$)-Au/Ge(111) is revealed by fully three-dimensional spin-resolved photoemission, as well as by density functional calculations. The large hexagonal Fermi surface, generated by the Au atoms, shows a significant splitting due to spin-orbit interactions. The planar components of the spin exhibit helical character, accompanied by a strong out-of-plane spin component with alternating signs along the six Fermi surface sections. Moreover, in-plane spin rotations towards a radial direction are observed close to the hexagon corners. Such a threefold-symmetric spin pattern is not described by the conventional Rashba model. Instead, it reveals an interplay with Dresselhaus-like spin-orbit effects as a result of the crystalline anisotropies.

Breaking the translational symmetry at the solid-vacuum interface strongly affects electrons, including their spin properties. The resulting structure-inversion asymmetry can induce a splitting of the surface bands based on the spin-orbit interaction, known as the *Rashba* effect [1]. The resulting lift of spin degeneracy is the basis for the emerging and important field of spintronics. As a current and prominent research object, topological insulators with their characteristic Dirac surface states show a spin-momentum locking [2]. Additional examples are surfaces formed by, or decorated with heavy atoms. This can be probed by angle-resolved photoelectron spectroscopy (ARPES), preferably with spin detection (SARPES), as shown, e.g., for surfaces like Au(111), Bi(111) [3,4] or surface alloys such as Bi/Ag(111) [5,6].

The realization of a strong Rashba effect in a metallic two-dimensional (2D) electron system at a semiconductor surface or interface would be particularly desirable, since it offers the perspective to manipulate spins electronically [7]. Different concepts to achieve spin filtering in Rashba systems, via ferromagnetic top electrodes [8] or by resonant tunneling [9], are intensively discussed. Studies in heterostructures show that the Rashba coupling strength can effectively be controlled by a gate field [10]. At semiconductor surfaces, it was demonstrated that heavy atoms may induce a particularly large Rashba splitting, as reported for the insulating bands formed by Bi and Tl reconstructions on Si(111) [11,12,13]. However, *conducting* spin-split states are needed to utilize spin control in electronic transport applications. To date, these have only been found in the β-phase of ($\sqrt{3}\times\sqrt{3}$)-Pb/Ge(111) [14], where a Rashba situation was observed using SARPES with in-plane spin detection only. In this regard, the Au-induced ($\sqrt{3}\times\sqrt{3}$)-reconstructed surface of Ge(111) with its spin-split *metallic* states represents a promising 2D electron system to be scrutinized in this study [15,16].

Additional perspectives are provided by the exploitation of the three-dimensional (3D) orientation of the spin vector at the Fermi surface (FS). For Tl/Si(111) and in surface alloys indications have been found that spins rotate out of plane along certain high symmetry directions [12,6]. These deviations of the spin vector from the ideal Rashba configuration (i.e., planar and tangential to the FS) may result from the interplay of Rashba and Dresselhaus-like spin-orbit terms in the Hamiltonian [17,18,19]. The concepts of a non-ballistic spin-field-effect transistor [20] or the electronic spin manipulation [21] based on such an interplay in 2D systems have been theoretically proposed and intensively discussed [22].

Recently, it was shown that the warped FS of the topological insulator $Bi_2Te_3$ is governed by an even more complex undulating spin structure [23,24,25]. However, no direct experimental demonstration by SARPES of such a behavior exists to date. Thus, a profound analysis of the 3D spin pattern for metallic surfaces is highly desirable. A key finding of our work is indeed that the 3D spin texture of the ($\sqrt{3}\times\sqrt{3}$)-Au/Ge(111) system displays an amazing similarity to that theoretically predicted in the recent studies of topological insulators [24,25]. Significant radial in-plane spin rotations are experimentally observed for the first time. Altogether, the undulating 3D spin pattern calls for an extension of the conventional Rashba model to include Dresselhaus-like contributions, resulting from the complex anisotropic potential landscape.

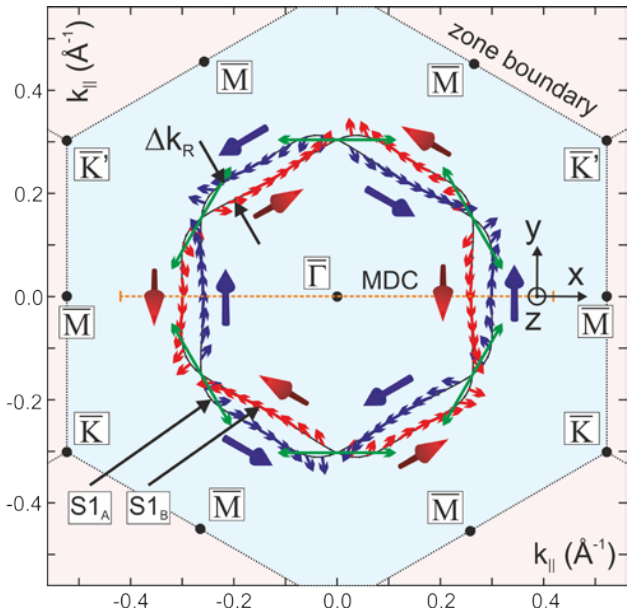

FIG. 1. (Color online) DFT Fermi surface (FS) in the 1$^{st}$ SBZ, exhibiting a surface state split into two bands (S1$_A$, S1$_B$) due to the spin-orbit interaction. Small arrows along the FS and their lengths indicate the in-plane spin component. For red (blue) color, the spin is rotated by up to 75° out of (into) the surface, highlighted by additional large arrows. Green color indicates fully in-plane alignment. The magnitude $\Delta k_R$ of the Rashba splitting between S1$_A$ and S1$_B$ varies with position along the FS. For experimental scans (e.g. dashed line), we use a local coordinate system (x, y, z).

An overview of the electron band and spin situation is provided by the FS topology obtained from DFT calculations in Fig. 1. The calculations have been performed within the local-density approximation (LDA), based on the commonly accepted conjugated honeycomb chained-trimer (CHCT) structure model [26,27]. The 21-layer slab calculation includes both spin-orbit interaction and a self-interaction correction (SIC) for the Au 5d states [15,28]. The SIC turned out to be crucial for the agreement between theoretical and measured spectra as has been demonstrated in Ref. 15. The calculated FS reproduces the *hexagonal* contour observed experimentally, originating from the Au-derived surface state S1 [15,29].

The DFT results in Fig. 1 predict a particular spin dependence of the FS: **i)** The FS displays a large *spin-splitting* into two bands S1$_A$ and S1$_B$ of $\Delta k_R \sim 0.04$ Å$^{-1}$, apart from a near degeneracy at the six corners of the hexagon. **ii)** The planar spin components show an in-plane helicity along the FS contour, suggestive of a Rashba scenario. **iii)** At the six hexagon corners of the FS the spin is fully aligned in-plane and perpendicular to the momentum vector, however, elsewhere along the FS it turns out of plane by up to 75°. **iv)** The z-component of the spin undulates between positive and negative values along the FS contour for each band in agreement with the $C_{3v}$ symmetry of the system. **v)** The in-plane spin components show a significant rotation towards a more radial orientation in between high symmetry directions.

Experimentally, SARPES has been performed at the Swiss Light Source using the COPHEE endstation [30,31], containing two Mott polarimeters for spin detection in 3D. The samples were prepared *in situ* by epitaxy of one monolayer of Au onto n-doped Ge(111) substrates, followed by annealing. The formation of the ($\sqrt{3} \times \sqrt{3}$)-reconstruction was verified by low energy electron diffraction. The spin texture is obtained in recording spin-resolved momentum distribution curves (MDCs), cutting the relevant band at 70 meV binding energy (increased statistics compared to the Fermi level).

The spin-resolved intensities $I_\alpha^\uparrow$ and $I_\alpha^\downarrow$ ($\alpha = x, y, z$; see local coordinate system in Fig. 1) along the positive (spin up) and negative (spin down) direction of the relevant detector channel, derived from an MDC in the 2$^{nd}$ surface Brillouin zone (SBZ) (along $\overline{M}-\overline{\Gamma}-\overline{M}$ outlined in Fig. 1), are plotted in Fig. 2(a) to (c). One finds that there is hardly any asymmetry for the x-component of the spin, i.e., $I_x^\uparrow$ and $I_x^\downarrow$ almost coincide. In turn, sizea-ble asymmetries exist for the y- and z-component, the latter being the largest. Besides the clearly resolved band splitting, another major fact becomes evident: The z-spin orientations on either side of the $\overline{\Gamma}$-point are *anti-symmetric* to each other. This is consistent with the spin signature predicted by our DFT calculations, and reflects the absence of time-reversal symmetry breaking.

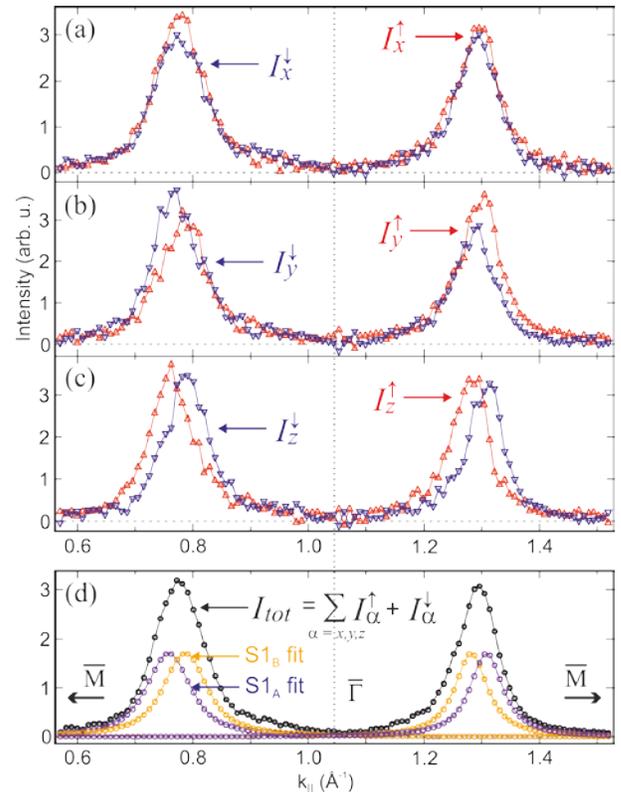

FIG. 2. (Color online) (a) Spin resolved intensities derived from scattering asymmetries for the x-component of the spin from an MDC in the 2$^{nd}$ SBZ along $\overline{M}-\overline{\Gamma}-\overline{M}$. Upward (downward) pointing triangles correspond to the spin orientation in positive (negative) direction of x. (b) Scattering asymmetries for the y-direction, and (c) the z-direction. (d) Total intensity MDC. The peak-doublets below originate from a two-step fitting routine [30] using the data in (a) to (c). Their maxima represent the spin-split band positions of S1$_A$ and S1$_B$.

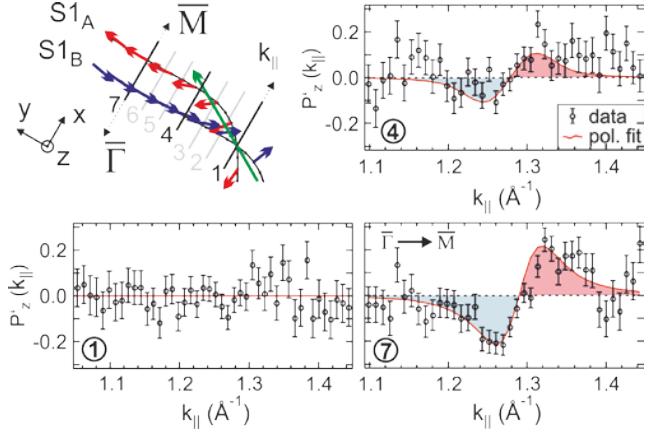

**FIG. 3**. (Color online) Measured MDC spin-polarization for the z-component as function of $k_{//}$ along three different scan positions of the FS in the 2$^{nd}$ SBZ. The red curves are fits to the data. The coordinate axes x, y, and z are strictly bound to the scan position.

The SARPES data allow further to determine the position and shape of each spin-band by applying a two-step fitting routine [30] to the total intensity MDC $I_{tot}$ in Fig. 2(d), which also involves the scattering asymmetries. The band-doublets in Fig. 2(d) result from that fit and, thus, clearly validate the existence of two separate and non-degenerate bands S1$_A$ and S1$_B$, being hidden in the spin-integrated data.

Moreover, we have carefully scanned the FS contours in seven steps (Fig. 3) for a 3D information on the spin-polarization

$$\vec{P}'(k_{||}) = \left(P'_x(k_{//}), P'_y(k_{//}), P'_z(k_{//})\right). \quad (1)$$

with $|\vec{P}'| \leq 1$. Each polarization component

$$P'_\alpha(k_{||}) = \frac{I_\alpha^\uparrow(k_{||}) - I_\alpha^\downarrow(k_{||})}{I_\alpha^\uparrow(k_{||}) + I_\alpha^\downarrow(k_{||})} \quad (2)$$

is extracted from the corresponding spin resolved intensities ($\alpha = x, y, z$). Regarding $P'_z(k_{||})$, the value is effectively zero at the scan position (SP) 1, and increases from there to SP 7. We also find that the z-spin-polarization is fully anti-symmetric with regard to the $\overline{\Gamma} - \overline{K}$ azimuth. This agrees well with the DFT prediction of two non-degenerate spin-polarized bands S1$_A$ and S1$_B$.

For better comparison with theory the measured MDC polarization $\vec{P}'(k_{||})$ has to be corrected for an incoherent unpolarized background and decomposed into the individual contributions of bands S1$_A$ and S1$_B$. For this purpose, we employed the two-step fitting routine introduced in Ref. 30, resulting in normalized band-specific spin polarizations $\vec{P} = (P_x, P_y, P_z)$ with $|\vec{P}|=1$. As an example Fig. 3 shows a fit to the measured z-polarization data. The resulting polarization of each individual state in comparison with the DFT values is summarized with regard to the k-space SP (1, 2,… 7) in Fig. 4. Turning to the z-component in Fig. 4(a), one finds that $P_z$ is almost constant along the scan area and exhibits a rather high value (mostly above 0.77) for both S1$_A$ and S1$_B$. It culminates at SP 7 with a value of 0.94, yet collapses to zero at SP1. This breakdown is also seen in the DFT results. It is locally so confined that the slope from high values to zero is not resolvable in SARPES. Here, averaging over a certain window in k-space is inevitable.

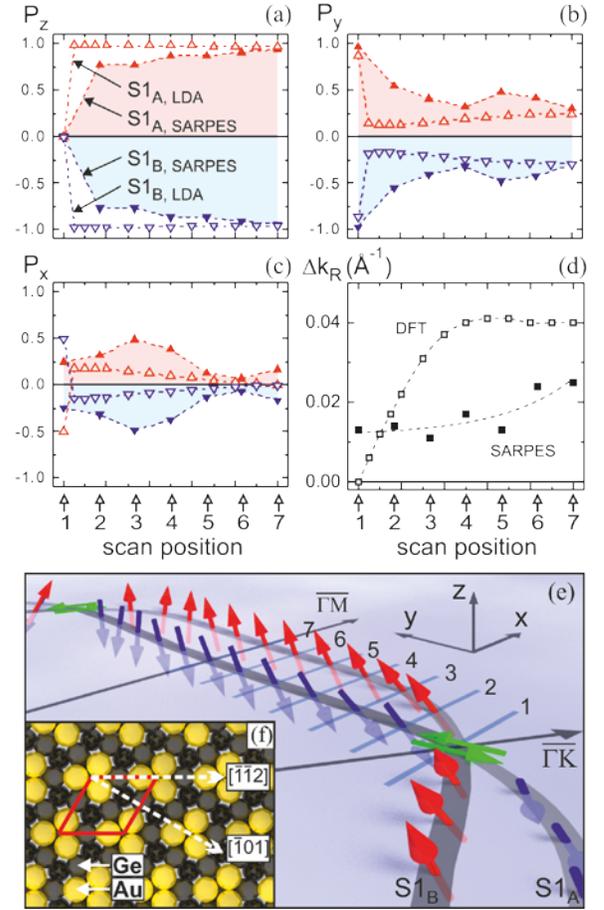

**FIG. 4**. (Color online) (a) – (c) z-, y-, and x-components ($P_z$, $P_y$, $P_x$) of the spin-polarization vector for each individual state obtained from both a fit of the SARPES data and DFT as a function of the SP 1, 2, … 7. (d) Rashba splitting $\Delta k_R$ as function of the corresponding SP. (e) 3D view of the spin-vector orientation as determined by SARPES. The SPs are the same as in Fig. 3. (f) CHCT model of ($\sqrt{3} \times \sqrt{3}$)-Au/Ge(111) with surface Brillouin zone and high-symmetry directions as indicated.

Another detail of the spin-vector rotation becomes evident when looking at the y- and x- components, as plotted in Fig. 4(b) and (c). In SARPES starting from SP 7, the in-plane y-component ("Rashba-type") gradually increases at the expense of $P_z$ until it assumes a value of order 1 at SP 1. The other in-plane component $P_x$, allowing for some scatter in the data, remains low within a range of $P_x \sim 0 - 0.5$. These observations are qualitatively reproduced in the DFT calculations.

The rotation of the spin-vector from out-of-plane towards in-plane is in good agreement between theory and experiment. It shows that the Rashba-type behavior (with surface-parallel spin orientation) is dominant only in $\overline{\Gamma} - \overline{K}$ azimuth. In contrast, dramatic deviations from this scenario are found along the FS contour in between, most prominently around the $\overline{\Gamma} - \overline{M}$ direction. Moreover, in close vicinity to the $\overline{\Gamma} - \overline{K}$ azimuth the deviation from helical to radial in-plane spin alignment is in good agreement with the predictions from the DFT. The resulting experimental situation is best visualized as a 3D plot of the spin-vector orientation along the FS in Fig. 4(e). Importantly, when viewing two adjacent straight FS segments of a split band ($S1_A$ or $S1_B$), the spin-vector points *upward* along one and *downward* for the other band (red and blue arrows in Fig. 4(e)). When moving to the next hexagon sector, the situation is reversed. In total, this results in an *undulating* change of orientation with threefold symmetry.

The band splitting $\Delta k_R$ obtained from experiment decreases along the rather straight section of the FS from SP7 to SP 1, see Fig. 4(d). At SP 1, the splitting is reduced to only half of its initial value. In comparing this to the DFT prediction, the calculated splitting is larger and rather constant in the vicinity of SP 7. However, in approaching SP 1 it drops gradually to almost zero.

Evidently, the spin orientations of $S1_A$ and $S1_B$ strongly depend on the $C_{3v}$ symmetry of the surface. High symmetry-directions in real space in the CHCT model of Fig. 4(f), directly map onto the behavior seen in k-space, here. The structural model is mirror-symmetric with respect to a $[\overline{1}\,\overline{1}2]$ line, which corresponds to the $\overline{\Gamma} - \overline{K}$ direction in reciprocal space. In contrast, along $[\overline{1}01]$ corresponding to $\overline{\Gamma} - \overline{M}$ in k-space, there is no mirror-plane symmetry. For non-degenerate states in a system with time-reversal symmetry, the spin direction must be perpendicular to the mirror plane. This explains the *in-plane* (and *perpendicular* to the *k*-vector) orientation of spins at the six $\overline{\Gamma} - \overline{K}$ directions. On the other hand, it follows from the combination of the time-reversal symmetry with the $\overline{\Gamma} - \overline{K}$ mirror-plane symmetry, that spin directions for non-degenerate states at the $\overline{\Gamma} - \overline{M}$ lines should be perpendicular to these lines, however, with a possible *non-zero z-component* of the spin. Due to the same symmetry arguments, the out-of-plane spin orientation of $S1_A$ ($S1_B$) alternates between two adjacent parallel sheets on both sides of the $\overline{\Gamma} - \overline{K}$ mirror plane.

Interestingly, both our experimentally and computationally observed spin pattern is highly reminiscent of the very recent theoretical prediction of the spin structure at the Dirac surface state of the topological insulator $Bi_2Te_3$ [25]. An up to fifth-order in *k* model Hamiltonian of $C_{3v}$ symmetry was proposed, which describes well the hexagonal warping of the Dirac state observed experimentally [32]. In addition, it reproduces the calculated *ab initio* spin structure along the Dirac-fermion band. Here, we adopt this model, though adding one modification. In the original model the *hexagonal warping* of the energy contours originates from the spin-orbit terms in the Hamiltonian. In contrast, as our DFT calculations show, the hexagonal shape of the Fermi contour is in first place an effect of the *crystal symmetry* alone, i.e., it is present already in the spinless part of the Hamiltonian. In order to reproduce our first-principles calculations shown in Fig. 1 in both cases, with and without the spin-orbit coupling, we add a sixth-order term to the spinless part of the model Hamiltonian. Finally, it takes the following form (see supplemental material [33]):

$$H(k) = \left(\frac{\hbar^2 k^2}{2m^*} - C + c_h(k_+^6 + k_-^6)\right)\sigma_0 + v(k_x\sigma_y - k_y\sigma_x) \quad (3)$$
$$+ \lambda(k_+^3 + k_-^3)\sigma_z + i\zeta(k_+^5\sigma_+ - k_-^5\sigma_-)$$

Unlike the case of $Bi_2Te_3$, Eq. (3) does not describe the surface band along its whole dispersion, but only reproduces the shape of the energy contour and the spin structure of the Au-derived surface bands close to the Fermi energy. By fitting Eq. (3) to the *ab-initio* results in Fig. 1, one obtains good agreement, with fit parameters as given in the supplemental material [33]. In this way, we are able to model the complex spin pattern with high accuracy.

To our knowledge, the present data are the first direct observation of a radial in-plane spin vector rotation at a semiconductor surface, accompanied by a large undulating perpendicular spin component. This complex spin configuration results from the special properties of the Au/Ge(111) interface: The Au atoms are closely embedded into the topmost Ge layer which leads to a significant hybridization of Au and Ge orbitals. This also results in pronounced potential gradients parallel to the surface in addition to the out-of-plane gradient, leading to the observed complex spin pattern.

In summary, our study unveils a highly spin-polarized FS at the Au/Ge(111) surface. A rather complex 3D undulating spin texture obeying the surface symmetry is found. This finding implies substantial modifications to a conventional Rashba picture, which is accurately modeled by introducing higher-order Dresselhaus terms to the Hamiltonian. The presence of a strong perpendicular spin component and deviations in radial direction from a Rashba-like helical spin structure point at the complexity of the potential landscape in "real-world" material systems. This has bearing for the discussion of a variety of topical solid-state themes, ranging from topological insulators and quantum spin-Hall systems to practical issues of spin manipulation in spintronics.

The authors acknowledge fruitful discussions with B. Trauzettel. The calculations have been performed at the Jülich Supercomputer Center. This work was supported by the DFG (FOR 1162) and the EU Framework Program FP7/2007-2013 (grant 226716).